\documentclass[10pt,journal,compsoc]{IEEEtran}

\usepackage[colorlinks=true,linkcolor=black,anchorcolor=black,citecolor=black,filecolor=black,menucolor=black,runcolor=black,urlcolor=black]{hyperref}

\usepackage[normalem]{ulem}
\usepackage[usenames, dvipsnames]{color}
\usepackage{xcolor}
\usepackage{booktabs}
\usepackage{todonotes}
\usepackage{url}
\usepackage{algorithm}
\usepackage{algpseudocode}
\usepackage{amsmath}
\usepackage[nocompress]{cite}
\usepackage{url}
\usepackage{hyperref}
\usepackage{xcolor}
\usepackage{alltt}

\definecolor{ChangedContentColor}{HTML}{009900} 

\definecolor{AddContentColor}{HTML}{C7372F} 

\definecolor{DeleteColor}{HTML}{808080} 

\definecolor{MoveColor}{HTML}{6495ED} 

\definecolor{FixColor}{HTML}{C7372F} 

\newcommand{\add}[1]{{#1}}

\newcommand{\SuppOne}[0]{Supp.~\S1}
\newcommand{\SuppTwo}[0]{Supp.~\S2}

\newcommand{\SuppFour}[0]{Supp.~\S4}
\newcommand{\SuppFive}[0]{Supp.~\S5}

\newcommand{\scope}[1]{\texttt{#1}}
\newcommand{\action}[1]{\texttt{#1}}
\newcommand{\target}[1]{\texttt{#1}}

\newcommand{\triplet}[1]{\textsc{#1}}
\newcommand{\task}[1]{\textit{``#1''}}

\begin{document}

\title{Visualization Tasks for Unlabeled Graphs}

\author{
Matt I. B. Oddo,
Ryan Smith,
Stephen Kobourov,
Tamara Munzner

  
  \IEEEcompsocitemizethanks{
    \IEEEcompsocthanksitem Matt Oddo, Ryan Smith, and Tamara Munzner are with the Department of Computer Science, University of British Columbia. Email: \{bofarull,rsmith0914,tmm\}@cs.ubc.ca 
    
    \IEEEcompsocthanksitem Stephen Kobourov is with the Department of Computer Science at the Technical University of Munich. Email: stephen.kobourov@tum.de \protect\\}
  \thanks{Manuscript received Month XX, 202X; revised XXX.}
}

\markboth{IEEE TRANSACTIONS ON VISUALIZATION AND COMPUTER GRAPHICS, ~Vol.~xx, No.~x, xxxx~xxxx}%
{Author \MakeLowercase{Oddo \textit{et al.}}: Visualization Tasks for Unlabeled Graphs}

\IEEEtitleabstractindextext{%
\begin{abstract}

We investigate tasks that can be accomplished with unlabeled graphs, which are graphs with nodes that do not have persistent or semantically meaningful labels \add{attached}. New visualization techniques to represent unlabeled graphs have been proposed, but more understanding of unlabeled graph tasks is required before these techniques can be adequately evaluated. Some \add{network visualization} tasks apply to both labeled and unlabeled graphs, but many do not translate between these contexts. We propose a data abstraction model that distinguishes the Unlabeled context from the increasingly semantically rich Labeled, Attributed, and Augmented contexts. We filter tasks collected and gleaned from the literature according to our data abstraction and analyze the surfaced tasks, leading to a taxonomy of abstract tasks for unlabeled graphs. \add{Our task taxonomy is organized according to the Target data under consideration, the Action intended by the user, and the Scope of the data at play.} We show the descriptive power of this task abstraction by connecting to concrete examples from previous frameworks, and \add{connecting} these abstractions to real-world problems. To showcase the evaluative power of the taxonomy, we perform a preliminary assessment \add{across 6 different network visualization idioms} for each task. For each combination of task and visual encoding, we consider the effort required from viewers, the likelihood of task success, and how both factors vary between small-scale and large-scale graphs.

\textbf{Availability:} Supplemental materials are available at \textit{\href{https://osf.io/e23mr}{osf.io/e23mr}}.

\end{abstract}

\begin{IEEEkeywords}
Task taxonomy, unlabeled graphs, invariant plots
\end{IEEEkeywords}}

\maketitle

\IEEEdisplaynontitleabstractindextext

\IEEEraisesectionheading{\section{Introduction}\label{Sec:Introduction}}

\IEEEPARstart{T}{ask} taxonomies support the analysis, design, and evaluation of visualization tools~\cite{Brehmer2013}. In the case of graph-structured data, where nodes are interconnected through edges, tasks may pertain to the topological structure of that connectivity, or attributes attached to nodes or edges~\cite{Lee2006}. In nearly all previous visualization contexts, graphs are assumed to be labeled graphs; that is, they have persistent unique identifiers for the nodes, even when the graph has no other attributes. 

However, there are network visualization approaches that do not make this assumption, and rely purely on topological information that is completely independent of node labels. These \emph{unlabeled graph} methods emphasize properties of graph data structures that remain unchanged when node labels are \add{locally reassigned, globally shuffled, or even ignored.} These techniques have been called \emph{invariant descriptor visualization} idioms~\cite{Oddo2024}, which we shorten here to \textbf{invariant plots}. Examples include the Network Portrait~\cite{Bagrow2008}, \add{GraphPrism~\cite{Kairam2012},} Graph Thumbnails~\cite{Yoghourdjian2016}, and Census plots~\cite{Oddo2024}. \add{A crucial property of invariant plots is that the visual patterns~\cite{Shu2024} they surface are always the same for structurally identical (isomorphic) graphs~\cite{Oddo2024}.}

Researchers may find it difficult to persuasively evaluate the efficacy and utility of invariant plots until a task taxonomy that pertains to them is introduced. Enumerating and categorizing what tasks can be performed on unlabeled graph data is an open problem that needs \add{addressing} before human-subjects studies can be run. The type of tasks supported by these invariant plots are fundamentally different from those targeted by the traditional views dedicated to semantically meaningful and persistent labels: while some tasks apply to both labeled and unlabeled graphs, many tasks designed for the labeled context do not translate to the unlabeled context. For example, the task of following the spread of disease between people through a contagion network presupposes persistent node labels to identify those specific people~\cite{Newman2003}. Conversely, some tasks that are challenging with traditional graph visualization approaches become feasible with visual encodings of unlabeled graphs. For example, pairwise comparison between two graphs could be easier with invariant plots that exploit a shared absolute coordinate system, than with node-link layouts where the spatial position of node points is arbitrary and unconstrained~\cite{Oddo2024}.

We attack this problem through a combination of literature review and reflective synthesis (Sec.~\ref{Sec:Process}). We identify which of the many concrete tasks discussed in previous work \add{are \emph{labels-not-required tasks}; that is, tasks for the unlabeled graph context}. We consider both abstract tasks stated explicitly in the visualization literature and implicit tasks that we infer from theoretical works, practitioner surveys, and domain-specific literature. To make these choices, we create a data abstraction that assigns graphs into the categories of Augmented, Attributed, Labeled, and Unlabeled (Sec.~\ref{Sec:DataAbstraction}), which allows us to filter tasks from existing literature (Sec.~\ref{Sec:RelatedWork}). We then propose our task taxonomy, which is an abstraction that maps concrete \add{labels-not-required tasks} into \add{\triplet{Target} + \triplet{Action} + \triplet{Scope}} triplets (Sec.~\ref{Sec:TaskTaxonomy}).

We show the generative power of our taxonomy by proposing new concrete tasks from the combinatorial possibilities of these triplets (Sec.~\ref{Sec:GeneratedTasks} and~\ref{Sec:ComplexTasks}). We also validate our task taxonomy through consideration of its descriptive, generative, and evaluative power~\cite{BeaudouinLafon2004}. We demonstrate its descriptive power by mapping many previously proposed concrete tasks into these abstract triplets; and provide some preliminary evidence of the ecological validity of the concrete tasks by connecting them to real-world problems stated in domain literature (Sec.~\ref{Sec:EcologicalValidity}). To showcase the descriptive and evaluative power of our taxonomy, we provide an extended example of its use in a preliminary assessment of 6 visual encodings, both traditional and invariant \add{plot} views, in relation to 17 concrete tasks (Sec.~\ref{Sec:ExampleAnalysis}). In this exercise, we made educated guesses about the effort demanded from viewers and the likelihood of success for both small and large graphs. This assessment is not intended to provide final answers to these questions, but rather to showcase the utility of the taxonomy for the design of future user studies.

In summary, the primary contribution of this paper is our \add{\triplet{Target} + \triplet{Action} + \triplet{Scope}} taxonomy of abstract visualization tasks for unlabeled graphs (Sec.~\ref{Sec:TaskTaxonomy}); the data abstraction model behind the construction of this task taxonomy is a secondary contribution (Sec.~\ref{Sec:DataAbstraction}).



\begin{figure*}[!t]
\centering

\fontsize{6.5}{8.2}\selectfont

\begin{tabular}{lp{11.2cm}cccc}

\textbf{First Author} & \textbf{Title} & \textbf{Tasks} & \textbf{Taxonomy} & \textbf{Year} & \textbf{Ref} \\
\hline

& & & & & \\

Ghoniem  & A comparison of the readability of graphs using node-link and matrix-based representations & Choose & X & 2004 & \cite{Ghoniem2004} \\

Lee      & Task taxonomy for graph visualization & Choose & X & 2006 & \cite{Lee2006} \\

Brehmer  & A multi-level typology of abstract visualization tasks & Choose & X & 2013 & \cite{Brehmer2013} \\

Munzner  & Visualization analysis and design & Choose & X & 2014 & \cite{Munzner2014} \\
Pretorius & Tasks for multivariate network analysis & Choose & X & 2014 & \cite{Pretorius2014} \\

Kerracher & A task taxonomy for temporal graph visualisation & Choose & X & 2015 & \cite{Kerracher2015} \\

\add{Murray}   & \add{A taxonomy of visualization tasks for the analysis of biological pathway data} & \add{Choose} & \add{X} & \add{2016} & \add{\cite{Murray2016}} \\

Pandey & A state-of-the-art survey of tasks for tree design and evaluation with a curated task dataset & Choose & X & 2022 & \cite{Pandey2022} \\

Filipov  & Are we there yet? A roadmap of network visualization from surveys to task taxonomies & Choose & X & 2023 & \cite{Filipov2023} \\

Bae      & Bridging network science and vision science: Mapping perceptual mechanisms to network visualization tasks & Choose & X & 2025 & \cite{Bae2025} \\

Bagrow   & Portraits of complex networks & Choose & & 2008 & \cite{Bagrow2008} \\

Kairam   & GraphPrism: Compact visualization of network structure & Choose & & 2012 & \cite{Kairam2012} \\

Yoghourdjian & Graph Thumbnails: Identifying and comparing multiple graphs at a glance & Choose & & 2016 & \cite{Yoghourdjian2016} \\

Bagrow   & An information-theoretic, all-scales approach to comparing networks & Choose & & 2019 & \cite{Bagrow2019} \\

Oddo    & The Census-Stub graph invariant descriptor & Choose & & 2024 & \cite{Oddo2024} \\

Herman  & Graph visualization and navigation in information visualization: A survey & Glean & & 2000 & \cite{Herman2000} \\

Newman  & The structure and function of complex networks & Glean & & 2003 & \cite{Newman2003} \\
Melancon & Just how dense are dense graphs in the real world? A methodological note & Glean & & 2006 & \cite{Melancon2006} \\
Shneiderman & Network visualization by semantic substrates & Glean & & 2006 & \cite{Shneiderman2006} \\
Hodas    & Friendship paradox redux: Your friends are more interesting than you & Glean & & 2013 & \cite{Hodas2013} \\
Kobourov & Force-directed drawing algorithms & Glean & & 2013 & \cite{Kobourov2013} \\
Perozzi  & DeepWalk: Online learning of social representations & Glean & & 2014 & \cite{Perozzi2014} \\

Behrisch & Matrix reordering methods for table and network visualization & Glean & & 2016 & \cite{Behrisch2016} \\

Grover   & node2vec: Scalable feature learning for networks & Glean & & 2016 & \cite{Grover2016} \\

Hebert-Dufresne & Multi-scale structure and topological anomaly detection via a new network statistic: The onion decomposition & Glean & & 2016 & \cite{HebertDufresne2016} \\

Ribeiro  & struc2vec: Learning node representations from structural identity & Glean & & 2017 & \cite{Ribeiro2017} \\

Tantardini & Comparing methods for comparing networks & Glean & & 2019 & \cite{Tantardini2019} \\

Pierri   & Topology comparison of Twitter diffusion networks effectively reveals misleading information & Glean & & 2020 & \cite{Pierri2020} \\

Sahu     & The ubiquity of large graphs and surprising challenges of graph processing: Extended survey & Glean & & 2020 & \cite{Sahu2020} \\

Kojaku   & residual2Vec: Debiasing graph embedding & Glean & & 2021 & \cite{Kojaku2021} \\
Lin      & Structural hole theory in social network analysis: A review & Glean & & 2022 & \cite{Lin2022} \\
Palowitch & GraphWorld: Fake graphs bring real insights for GNNs & Glean & & 2022 & \cite{Palowitch2022} \\
Wang     & Quantification of network structural dissimilarities based on network embedding & Glean & & 2022 & \cite{Wang2022} \\

Faskowitz & Connectome topology of mammalian brains and its relationship to taxonomy and phylogeny & Glean & & 2023 & \cite{Faskowitz2023} \\

Rosen    & Homology-preserving multi-scale graph skeletonization using mapper on graphs & Glean & & 2023 & \cite{Rosen2023} \\

Shvydun  & Models of similarity in complex networks & Glean & & 2023 & \cite{Shvydun2023} \\

Di Bartolomeo & Evaluating graph layout algorithms: A systematic review of methods and best practices & Glean & & 2024 & \cite{DiBartolomeo2024} \\

\add{Luppi}    & \add{Systematic evaluation of fMRI data-processing pipelines for consistent functional connectomics} & \add{Glean} & & \add{2024} & \add{\cite{Luppi2024}} \\

Puxeddu  & Relation of connectome topology to brain volume across 103 mammalian species & Glean & & 2024 & \cite{Puxeddu2024} \\

Shu      & Does this have a particular meaning? Interactive pattern explanation for network visualizations & Glean & & 2024 & \cite{Shu2024} \\

Velikonivtsev & Challenges of generating structurally diverse graphs & Glean & & 2024 & \cite{Velikonivtsev2024} \\

Wong     & Discovery of a structural class of antibiotics with explainable deep learning & Glean & & 2024 & \cite{Wong2024} \\

Ahn      & A task taxonomy for network evolution analysis & Exclude & X & 2014 & \cite{Ahn2014} \\

Kindlmann & An algebraic process for visualization design & Exclude & X & 2014 & \cite{Kindlmann2014} \\

Saket    & Group-level graph visualization taxonomy & Exclude & X & 2014 & \cite{Saket2014} \\

Vehlow   & The state of the art in visualizing group structures in graphs & Exclude & X & 2015 & \cite{Vehlow2015} \\

Beck     & A taxonomy and survey of dynamic graph visualization & Exclude & X & 2017 & \cite{Beck2017} \\

Nobre    & The state of the art in visualizing multivariate networks & Exclude & X & 2019 & \cite{Nobre2019} \\

Dimara  & A task-based taxonomy of cognitive biases for information visualization & Exclude & X & 2020 & \cite{Dimara2020} \\

McNutt  & Surfacing visualization mirages & Exclude & X & 2020 & \cite{McNutt2020} \\

Pretorius & Visual inspection of multivariate graphs & Exclude & & 2008 & \cite{Pretorius2008} \\

Malliaros & Clustering and community detection in directed networks: A survey & Exclude & & 2013 & \cite{Malliaros2013} \\

Angluin  & Effective storage capacity of labeled graphs & Exclude & & 2014 & \cite{Angluin2014} \\

Harenberg & Community detection in large-scale networks: a survey and empirical evaluation & Exclude & & 2014 & \cite{Harenberg2014} \\

Battaglia & Relational inductive biases, deep learning, and graph networks & Exclude & & 2018 & \cite{Battaglia2018} \\

Cai      & A comprehensive survey of graph embedding: Problems, techniques, and applications & Exclude & & 2018 & \cite{Cai2018} \\

Bertolero & On the nature of explanations offered by network science: A perspective from and for practicing neuroscientists & Exclude & & 2020 & \cite{Bertolero2020} \\

Garcia   & Exploiting symmetry in network analysis & Exclude & & 2020 & \cite{Garcia2020} \\

Dey      & Community detection in complex networks: From statistical foundations to data science applications & Exclude & & 2021 & \cite{Dey2021} \\

Chami    & Machine learning on graphs: A model and comprehensive taxonomy & Exclude & & 2022 & \cite{Chami2022} \\

Dvorak   & Radius, girth and minimum degree & Exclude & & 2022 & \cite{Dvorak2022} \\

He       & A survey of community detection in complex networks using nonnegative matrix factorization & Exclude & & 2022 & \cite{He2022} \\

Ye       & A comprehensive survey of graph neural networks for knowledge graphs & Exclude & & 2022 & \cite{Ye2022} \\

Jin      & A survey of community detection approaches: From statistical modeling to deep learning & Exclude & & 2023 & \cite{Jin2023} \\

Macocco  & Intrinsic dimension as a multi-scale summary statistics in network modeling & Exclude & & 2024 & \cite{Macocco2024} \\

\end{tabular}

\vspace{4pt}

\caption{We report on the 65 papers in our literature review, providing first author name and titles for each. The list is organized according to whether we chose tasks \add{directly}, gleaned implicitly mentioned tasks, or excluded all tasks from the paper (Tasks). We \add{also} document whether the paper features a task taxonomy (Taxonomy), publication (Year), and provide the citation (Ref).}

\vspace{-8pt}

\label{PapersTable}
\end{figure*}


\vspace{-0.1cm}

\section{Process}\label{Sec:Process}

We interleaved six process stages through multiple rounds of iterative refinement that occurred across six months, through individual work from the first author and weekly meetings with two other senior authors, until convergence where consensus was reached. Our process included the methods of literature review and reflective synthesis.

\vspace{-0.2cm}

\subsection{Collecting Concrete Tasks}

We use the term \emph{concrete tasks} to mean a prose version of a task, which may be stated using either domain-specific or abstracted language. To identify concrete tasks from previous work, the first author conducted a literature review using seed papers on network task taxonomies, visualization techniques, and surveys already known to the authors, from which a broader forward and backward citation search for relevant material followed. At this stage, 65 papers were identified. Task taxonomy papers included explicit abstract tasks that we could simply extract. For example, \task{is the node an articulation node?} in Lee et al.'s seminal taxonomy of network visualization tasks~\cite{Lee2006}. We also gleaned end-user tasks from non-taxonomy papers, either inferring them through definitions of graph-theoretic concepts and terminology, or through descriptions of the use cases for algorithms or techniques.

\vspace{-0.2cm}

\subsection{Developing Data Abstraction}

\begin{figure*}[!t]
\centering
\vspace{9pt}
\includegraphics[width=0.97\linewidth]{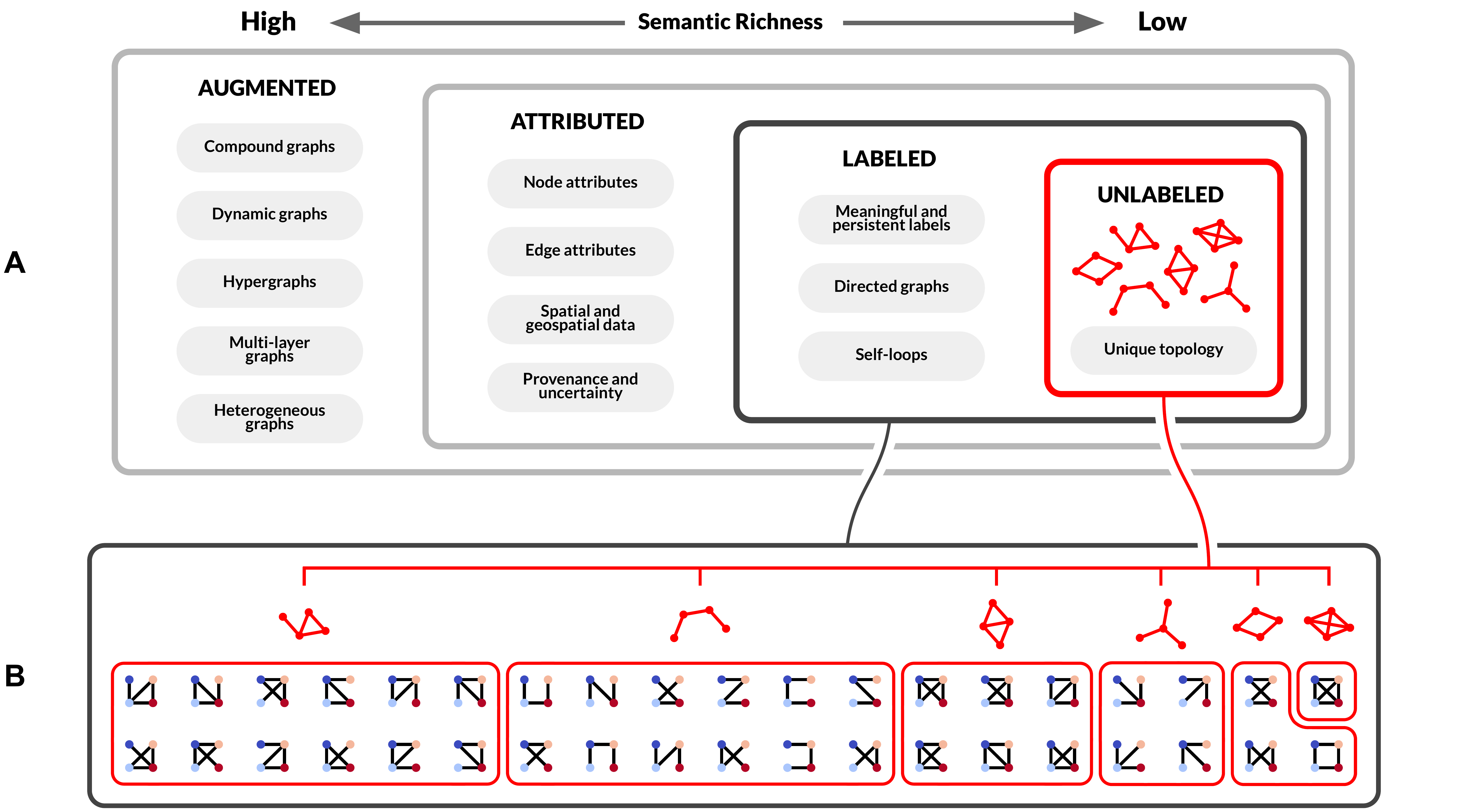}
\vspace{3pt}
\caption{Data abstraction model with an illustrative example. \textbf{(A)} In our data abstraction model, nested layers progressively decrease in semantic richness, shown here from high on the left to low on the right: \emph{Augmented} is a stand-in term for named graph models with rich associated semantics, in \emph{Attributed} non-topological semantic information is allowed to be associated with nodes or edges, in \emph{Labeled} there are persistent unique identifiers that can convey semantic meaning, and \emph{Unlabeled} is the innermost layer where all semantic information is stripped away, leaving only unique (non-isomorphic) unlabeled graphs. \textbf{(B)} As an illustrative explainer, drawn in red are the 6 unlabeled graphs of 4 nodes. If persistent labels (here shown as node colors) exist, as in the Labeled layer, this set of 6 graphs expands to the set of 38 labeled 4-node graphs. Within each of the 6 groupings based on graph topology (red enclosures) the graphs are structurally equivalent (isomorphic), and distinguishable from each other only through the meaning conveyed through the node labels.}
\vspace{-3pt}
\label{DataAbstraction}
\end{figure*}

As we considered which concrete tasks to include and which to exclude, we developed a data abstraction model through reflective synthesis that could serve as a filter to exclude tasks that cannot be undertaken with unlabeled graphs. It assigns graphs into four nested layers: Augmented, Attributed, Labeled, and Unlabeled, as shown in Fig.~\ref{DataAbstraction}.

\vspace{-0.2cm}

\subsection{Filtering by Data Abstraction}

We used our data abstraction model to filter the set of collected concrete tasks. We excluded as irrelevant any task that required information from the Augmented, Attributed, or Labeled data layers; we call such tasks \add{\textbf{labels-required tasks}}. We only kept tasks that pertained to the Unlabeled layer, which we call \add{\textbf{labels-not-required tasks}}.

This filtering based on our data abstraction allowed us to exclude all tasks from 24 of the papers. We chose to include at least one task from each of the other 41 papers (15 taxonomies, and 26 non-taxonomy papers). In most cases, these papers had a mix of \add{labels-required} and \add{labels-not-required} tasks. As with developing the data abstraction, our reflective synthesis process involved analysis by the first author, discussed in detail with the other senior authors in weekly meetings, and iterative refinement until a consensus was reached between all authors.

For example, the Pandey et al. survey on tasks for hierarchical visual encodings of tree topology~\cite{Pandey2022} contained many tasks that we excluded through data abstraction filtering by noting that they require attributes (e.g. semantic node label A, B, or C) and hierarchical wayfinding (e.g. hierarchical level Y, X, or Z from root node). To illustrate, the concrete tasks \task{is the tree balanced or un-balanced?} \add{and \task{count the total number of levels in the tree}} require an established root node, which is a semantic label associated with a specific node, and therefore a \add{labels-required task, which we exclude}. \add{In contrast, labels-not-required tasks such as \task{count all leaf nodes}} that appear in this survey we include.


Furthermore, we used our data abstraction to glean tasks from papers that are not taxonomy papers, for example Newman's landmark 2003 paper on complex network structure~\cite{Newman2003} implicitly suggests that \task{finding the length of the longest geodesic} and \task{counting the number of longest geodesics} are relevant tasks to draw insights about the network's maximum traversable distances, and the induced paths that support these traversals. Both of these tasks are \add{labels-not-required tasks}, because the length of the graph's diameter and the number of longest shortest paths are targets that can be found without the need of semantically meaningful or persistent node labels.

Moreover, our data abstraction allowed us to proactively rule out many potential sources of concrete tasks without explicitly collecting such tasks. For example, hybrid network layouts such as NodeTrix~\cite{Henry2007} are designed for grouping nodes according to attribute values and identifying specific nodes based on semantic labels; thus \add{they address labels-required tasks, which we exclude}.

Fig.~\ref{PapersTable} summarizes our results, \add{with papers coded by whether we Choose (15), Glean (27), or Exclude (23) tasks. Most tasks are gleaned so most papers we classify as Glean; but if a paper has a single Choose task, then we classify the paper as Choose.} We provide a \add{more detailed} spreadsheet version in~\add{\SuppOne}.


\vspace{-0.25cm}
\subsection{Mapping Task Abstraction}

We developed the task abstraction (Sec.~\ref{Sec:TaskTaxonomy}) that covered \add{our collected labels-not-required} concrete tasks with a concise set of terms through multiple rounds of reflective synthesis. We finalized the framework of three levels \add{(Target, Action, and Scope)} early on, but refined the set of items at each level extensively. As with previous process stages, our reflective synthesis process involved ideation from the first author, discussed in detail with the other senior authors in weekly meetings. Iterative refinement proceeded until a consensus was reached between all authors.

\vspace{-0.2cm}

\subsection{Analyzing Visual Encodings}

We considered the strengths and weaknesses of six different visual encodings for specific \add{labels-not-required} tasks using reflective synthesis, presented in Sec.~\ref{Sec:ExampleAnalysis}. We considered the encoding results across the set of 81 eclectic benchmark networks that we collected in previous work~\cite{Oddo2024}. We drew on the experience and expertise of the authors: two are doctoral students, and two are senior visualization researchers. The first author created 9-by-9 faceted panels featuring the 81 benchmark networks, one panel for each of the six visual encodings. The authors convened over weekly meetings to discuss task performance in relation to these visual encodings. These six panels are shown in\add{~\SuppFive}.

\subsection{Generating Concrete Tasks}

We generated new \add{labels-not-required} concrete tasks \add{from the} abstraction triplets where we had not found compelling examples articulated in prior work, discussed in Sec.~\ref{Sec:ConcreteTasks}. We again used reflective synthesis, spearheaded by the first author generating new tasks and the two senior authors discussing their applicability in weekly meetings.

\subsection{\add{Additional Concrete Tasks}}

\add{We compiled a list of 30 tasks (26 chosen, 4 gleaned) from the Choose set of 15 papers (Fig.~\ref{PapersTable}), and analyzed each task according to our \add{\triplet{Target} + \triplet{Action} + \triplet{Scope}} triples, with process described in previous stages above. We included at least one task from each paper, and no more than four from any individual paper. Full table available in~\SuppTwo.}


\vspace{-0.1cm}

\section{Data Abstraction}\label{Sec:DataAbstraction}

Our data abstraction model distinguishes four layers of semantic richness to describe networked data, as shown in Fig.~\ref{DataAbstraction}-A.

\vspace{-0.15cm}

\subsection{Model Layers}

We use the term \emph{concrete tasks} to mean a prose version of a task, which may be stated using either domain-specific or abstracted language.

\textbf{Augmented}. This outermost layer of our data abstraction (Fig.~\ref{DataAbstraction}-A) addresses the rich semantic complexity of real-world networked data; a stand-in for named graph models, which are formalisms with specific facets of associated semantic data that help us interpret and analyze the model. To illustrate, \emph{dynamic graphs} have node and edge attributes parsed as time slices~\cite{Ahn2014, Kerracher2015, Beck2017}, \emph{compound graphs} can have disjoint or overlapping groups assigned to nodes~\cite{Saket2014, Vehlow2015}, \emph{hypergraphs} require a generalized definition of relationship that allows more than two nodes to be connected by the same edge at once~\cite{Filipov2023}, and \emph{heterogeneous graphs} have multiple different kinds of attributes associated to nodes and edges~\cite{Pretorius2014, Nobre2019, Chami2022}.

Augmented graphs cannot be unlabeled graphs by definition: they explicitly associate semantic information beyond purely topological structure. Any task that pertains to the use of such formalism semantics can easily be ruled out as impossible to achieve with an unlabeled graph.

\textbf{Attributed}. Even graphs without the semantic complexity of formalisms can contain \emph{data attributes} assigned to nodes and edges, which encode domain-specific information that augments the topological structure. For example, in a social network, semantic data can include node attributes such as a person’s height and age~\cite{Newman2003}, while in a biological system such semantic data may take the form of edge attributes representing the reactivity of different metabolites~\cite{Murray2016}. Attribute data is used extensively in graph drawing~\cite{Kobourov2013}.

These attributes on nodes or edges are independent from and orthogonal to the topological structure of the network. Signed values (positive or negative) and scalar weights (for nodes and/or edges) are special cases of attributes. Any task that relies on the use of attributes can be ruled out as impossible to achieve with an unlabeled graph.

\textbf{Labeled}. The next lowest layer of semantic richness is graphs with \emph{labels}, which we define as semantically meaningful unique identifiers that are persistent and can serve to distinguish nodes from each other. Although this particular distinction is not common, we propose it because even graphs that are typically considered as being attribute-free in the visualization literature do in fact have labels.

For instance, in the context of Graph Neural Networks (GNNs), the term \add{\emph{non-attributed graphs}} refers to networked data that lacks data at the Attributed layer, but does have node identifiers~\cite{Cui2022}, which is data at the Labeled layer. Similarly, directed graphs require labels to establish persistent edge direction, which conversely implies that unlabeled graphs cannot be directed. By extension, unlabeled graphs cannot have self-loops.

Although labels could be considered as a special case of attributes, we distinguish them from the more general attributes in the previous Attributed layer because the Labeled layer is at the heart of the distinction between labeled and unlabeled graphs. Any task that relies on given labels is inapplicable to unlabeled graphs \add{(more in Sec.~\ref{Sec:TaskModelImplications})}.

\textbf{Unlabeled}. At the innermost layer, outlined in red in Fig.~\ref{DataAbstraction}-A, is the connectivity information of networked data, which establishes how nodes are connected via edges, or the graph's skeletal topology. In this layer, nodes can be distinguished from each other only in terms of topological connectivity, not through persistent names, so there is no semantic information at all at this layer. All graphs at the Unlabeled layer are topologically unique (non-isomorphic) to each other.

A specific example of a concrete task that cannot be undertaken in the Unlabeled context is \task{find the shortest path between node A and node B}\add{~\cite{Ghoniem2004}}. The labels of node A and node B are precisely the information not available. By extension, most pathfinding tasks are also impossible to execute in the Unlabeled context, because they require identifying specific nodes as the start or end of the induced path.

\vspace{-0.3cm}

\subsection{Implications\add{: Data Abstraction Model}}\label{Sec:DataAbstractionImplication}

To illustrate the difference between Labeled and Unlabeled layers, Fig.~\ref{DataAbstraction}-B compares labeled and unlabeled graphs: with 4 nodes there are only 6 different unlabeled graphs (node-link views in red), while there are 38 possible labeled graphs, which we show with node colors to indicate the labels, rather than the more typical approach of using text strings (which would be too small to read).

The correspondence with the unlabeled graphs is shown through the 6 groupings (red enclosures) in Fig.~\ref{DataAbstraction}-B, each of which has a respective unlabeled graph. Although the topological structure is the same within the groups, the labeled graphs are distinguishable from each other through the persistent labels that can convey meaning. For example, in the third red enclosure from the right, with 4 labeled graphs, the single degree-3 node can be bound to each of the different labeled cases, shown with the different colors. Without labels there is no difference between the four graphs within that group, so we have only a single unlabeled graph corresponding to each grouping. In more mathematical terms, the unlabeled graph is the \add{\textbf{isomorphism class}} of the labeled graph~\cite{Angluin2014}.

Within the Unlabeled layer, node identifiers are unique but non-persistent and arbitrary. However, to instantiate a graph in any computational context, such as to load networked data in memory or perform a hop-based traversal of it, there must be some kind of unique identifiers that would serve as labels during the analysis run time. To clarify, what we emphasize in this paper with our data abstraction model is not the presence of labels themselves, but whether the labels carry semantic meaning\add{; and how this meaning impacts algorithmic and visual idiom design (Sec.~\ref{Sec:TaskModelImplications})}.


\vspace{-0.25cm}
\section{Related Work}\label{Sec:RelatedWork}

We discuss previous work on general task taxonomies, network task taxonomies that pertain to the Unlabeled layer of our data abstraction, network task taxonomies that cover semantically richer data abstractions, and tasks from the network visualization technique literature.

\vspace{-0.25cm}
\subsection{General Task Taxonomies}

An early taxonomy from Shneiderman~\cite{Shneiderman1996} provides a high-level breakdown of tasks and of data types, while Amar et al.~\cite{Amar2005} provide a low-level taxonomy of analytic tasks. In 2013 Brehmer and Munzner~\cite{Brehmer2013} proposed bridging from low-level tasks to higher-level tasks with a multi-level typology of abstract tasks. Munzner~\cite{Munzner2014} breaks down tasks into action-target pairs; we extend this approach in our taxonomy.

From this datatype-agnostic foundation, many other taxonomies focus on narrower, more specialized scopes of specific data types. For example, Valiati et al.~\cite{Valiati2006} cover multidimensional data. Nusrat and Kobourov cover the intersection of geographic data and the visual encoding technique of cartograms~\cite{Nusrat2016}. None of these papers focus on the data type of networks, which is the subject of our work. 

\vspace{-0.25cm}
\subsection{Network Task Taxonomies}

The seminal network task taxonomy from Lee et al.~\cite{Lee2006} includes some topology-based tasks that are well-suited for unlabeled graphs, which we incorporate into our list of applicable concrete tasks. However, the majority of their topology tasks begin with a \task{given a set of nodes} as the starting point, which requires meaningful semantic labels that are not available with unlabeled graphs. Moreover, they do not cover the \scope{Multiple} or \scope{Pair} scopes at all \add{(more details about task taxonomy scopes in Sec.~\ref{Sec:TaskTaxonomyScope}).}

Although the taxonomy by Kerracher et al.~\cite{Kerracher2015} primarily focuses on temporal graphs, they also discuss the more general issue that structural tasks with no attribute metadata involved are an important category, many of which we include in our list. They identify comparison as a task of interest, corresponding to our \action{Compare} action at the \scope{Pair} scope. However, they do not address the \scope{Multiple} scope. 

Some structural comparison tasks relevant to unlabeled graphs are described by Bae et al.~\cite{Bae2025}, but the main focus of their paper is the perceptual mechanisms involved in network pattern detection. 

While some aspects of these taxonomies cover tasks that pertain to the Unlabeled data abstraction layer (Fig.~\ref{DataAbstraction}-A), their main focus is tasks that primarily rely on semantic data at higher levels of semantic richness, and none have comprehensive coverage of the full set of tasks that apply to unlabeled graphs.

\vspace{-0.25cm}
\subsection{Labeled Graph Task Taxonomies}

Many specialized task taxonomies address the types of graphs that we assign to the \add{Augmented} layer of our data abstraction, including dynamic graphs that focus on changes over time of Attributed networks~\cite{Ahn2014, Beck2017, Kerracher2015}, heterogeneous graphs with diverse multivariate attributes~\cite{Pretorius2014, Nobre2019}, and compound graphs with group semantics as node attributes~\cite{Saket2014, Vehlow2015}. Others focus on domain-specific analysis, such as biological pathway networks~\cite{Murray2016}, with tasks that rely on rich semantic attributes. However, none of these are \add{labels-not-required tasks, or tasks} applicable to unlabeled graphs. A comprehensive 2023 survey of network visualization task taxonomies~\cite{Filipov2023} illustrates how much of the previous work is dedicated to networks with \add{node and edge} attributes \add{attached}, confirming the need for a task taxonomy for the unlabeled graph context. 

We can also explore how real-world constraints shape the kinds of tasks researchers prioritize in practice. A practitioner survey by Sahu et al.~\cite{Sahu2020} highlights that large-scale network analysis often involves graphs exceeding a billion edges, with dynamic structures and complex label attributes.

\vspace{-0.25cm}
\subsection{Network Visualization Techniques}

Visualization technique papers often include explicit lists of tasks, despite not being framed as task taxonomies. We considered many such papers, and gleaned tasks from them that appeared to fit within the Unlabeled context. For example, the GraphPrism~\cite{Kairam2012} system is explicitly aimed at producing invariant plots, so its tasks were a good fit with our work. Some papers that are not aimed at invariant network visualization in particular nevertheless did serve as sources for concrete tasks, such as the Semantic Substrates~\cite{Shneiderman2006} technique. Many technique papers list only tasks that require additional data layers, so we did not include them. For example, systems focused on the application domain of social network analysis, such as MatLink~\cite{Henry2007}, rely on persistent semantic labels.


\section{Task Taxonomy}\label{Sec:TaskTaxonomy}

Our abstract task taxonomy for unlabeled graphs has three dimensions: each task is broken down into a \add{\triplet{Target}, an \triplet{Action}, and a \triplet{Scope}}, as shown in Fig.~\ref{TaskTaxonomy}-A. The \add{\triplet{Target} + \triplet{Action} pairs follow the Action-Target model of Munzner~\cite{Munzner2014}, but with flipped order: a \triplet{Target} is a noun specifying some aspect of the data that is of interest to the user; while an \triplet{Action} is a verb that defines user goals. In addition, the \triplet{Scope} is an ordered axis that captures the scale of the data, with the location within the axis inferred from the \triplet{Target}.} Our taxonomy includes targets that are well-known concepts from the graph theory literature, or can be precisely expressed in terms of topological structure. Given our focus on unlabeled graphs, no task can depend on persistent labels with semantic meaning, although hop-based traversal can be supported through non-persistent arbitrary labels.

\vspace{-0.2cm}
\subsection{Scope}\label{Sec:TaskTaxonomyScope}

The \triplet{Scope} dimension has 5 ordered levels, indicating the scale of the topological data under consideration. At the highest level of scope are \scope{Multiple} graph tasks, which require at least three graphs; followed by the \scope{Pair} scope, which covers exactly two graphs. Next, at the \scope{Single} scope the primary focus is on the topology of the whole graph; followed by the \scope{Subgraph} scope, with a focus on subregions. Finally, the \scope{Constituent} scope is the most atomic, focused on individual nodes and edges, the elementary constituents of graph topology.

In this paper, we use \emph{graph} to mean a single connected component, for simplicity of exposition. Analysis of a disconnected graph with multiple connected components could be handled in our framework by splitting it into multiple graphs.

\vspace{-0.2cm}
\subsection{Targets}\label{Sec:TaskTaxonomyTargets}

We identify semantic targets and assign them at their respective scope in our taxonomy. We discuss targets in order from \add{a broad scope}, which covers multiple graphs simultaneously; to \add{a narrow scope}, where analysis is at the level of nodes and edges, the elementary constituents of unlabeled graph topology. The list of targets that we present here is intended to be representative \add{and illustrative, but} not exhaustive. Adding new targets \add{to the enumeration} is consistent with the spirit of this taxonomy, as long as they are assigned to the appropriate scope.

At the \scope{Multiple} and \scope{Pair} scopes, the target is simply the \target{graph} itself; that is, the Unlabeled graph topology (Sec.~\ref{Sec:DataAbstraction}) of a simple graph consisting of a single connected component. At the \scope{Pair} scope there are two such targets, and at the \scope{Multiple} scope there are many. 

At the \scope{Single} taxonomy scope, the target pertains to the entire graph, encompassing all of its nodes and edges. Many of the targets are names from graph theory literature that describe global topological characteristics\add{, generally summarized as graph \target{type}. For example, a} graph of $N$ nodes is a \target{tree} when it has $N-1$ edges and a \target{complete} graph when it has ${N \choose 2}$ edges. While these extreme cases are well-defined, there exists a vast diversity of topologies between these fully-connected extremes. \add{Other graph \target{type} examples include}: \target{random}, where any two nodes are connected following an established probability, leading to a Poisson degree distribution~\cite{Newman2003}; \target{small-world}, where the topology is random with internal cycles that connect otherwise far apart nodes; \target{scale-free}, where a handful of nodes have exponentially large number of neighbors, while the majority of nodes only have a few, leading to an exponential degree distribution~\cite{Newman2003}; \target{regular}, where all nodes have the same or nearly the same number of neighbors; and \target{multipartite}, with 2 or more sets of nodes that are not connected within their set, but exclusively between sets. \add{There can also be} approximations; for example, a \target{near-regular} \add{percolated lattice}~\cite{Newman2003} or a \target{quasi-tree} are structurally noisy versions of their respective crisp topologies. This list of \add{\target{type}} examples is not a complete enumeration; rather, \add{we present a} broad representative \add{set}.


\begin{figure}[!t]
\vspace{1pt}
\centering
\includegraphics[width=0.92\linewidth]{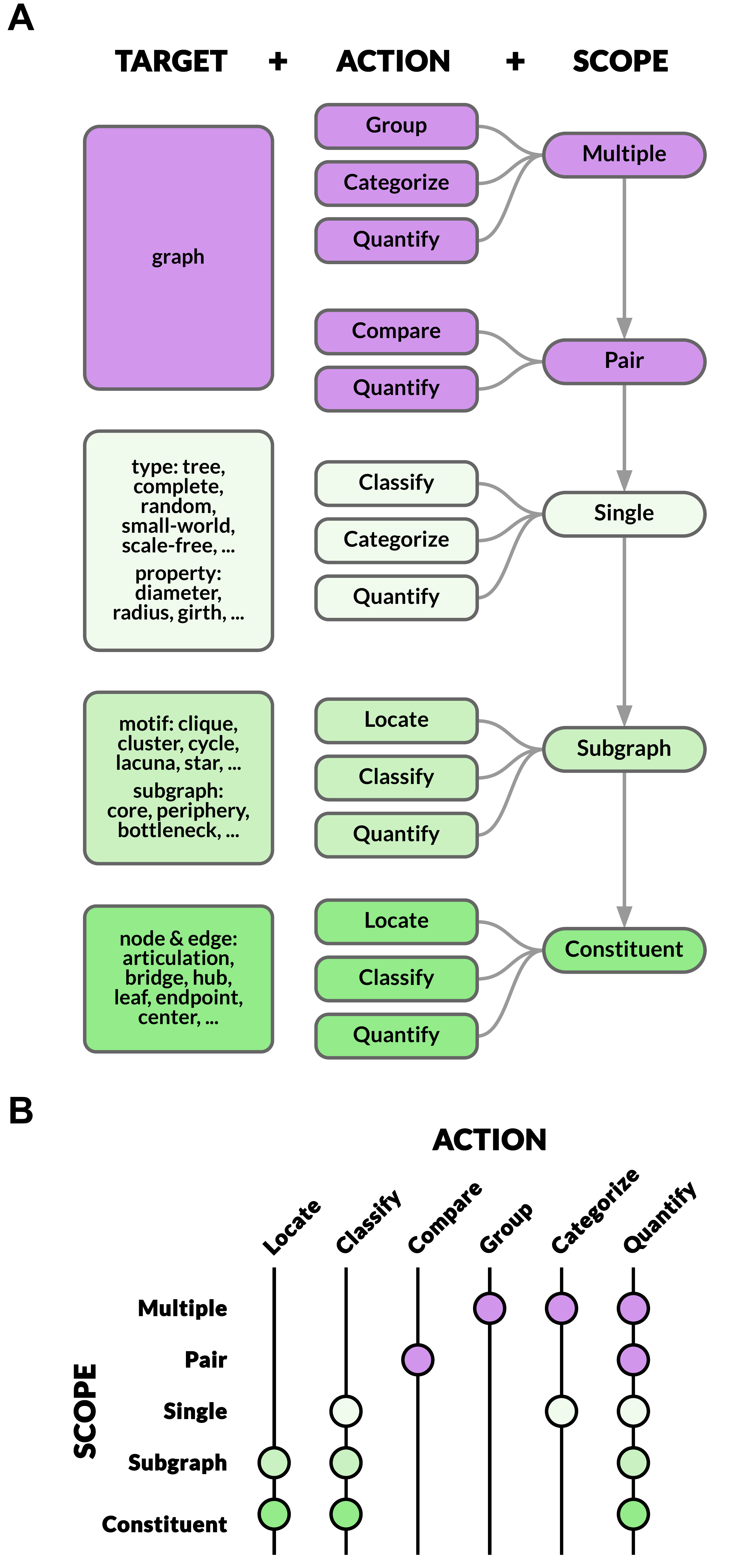}
\vspace{2pt}
\caption{Our task taxonomy has three dimensions: \add{\triplet{Target}}, \triplet{Action}, and \add{\triplet{Scope}}. \textbf{(A)} We can generate unlabeled graph tasks by combining an \triplet{Action} and a \triplet{Target} at a given \triplet{Scope}. In purple the broad-scope \scope{Multiple} and \scope{Pair} scopes, while in green the narrow-scope \scope{Single}, \scope{Subgraph}, and \scope{Constituent} scopes (the darker the green hue, the narrower the scope). Our \triplet{Target} list is illustrative not exhaustive, \add{as indicated by ellipses}. \textbf{(B)} Combinatorial space of \triplet{Scope} + \triplet{Action} pairs.}
\vspace{-10pt}
\label{TaskTaxonomy}
\end{figure}


\begin{figure}[!t]
\vspace{14pt}
\centering
\includegraphics[width=0.82\linewidth]{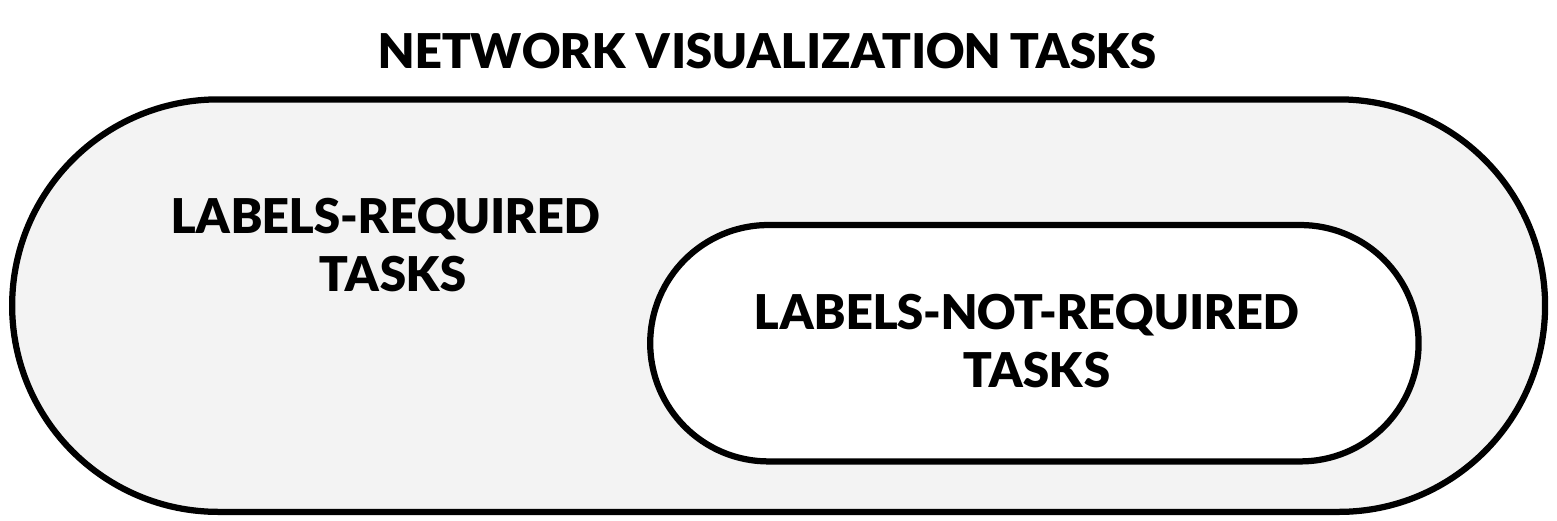}
\vspace{2pt}
\caption{\add{The set of labels-required tasks is the complement set, or the set that remains when the labels-not-required tasks are removed from the set of all network visualization tasks}.}
\vspace{-10pt}
\label{TaskSets}
\end{figure}


Targets at the \scope{Single} scope also include single-integer summary statistics\add{, or a \target{property},} that require global computation based on distances counted as hops through the graph topology. The \target{diameter} is the hop count of the longest shortest path, the \target{radius} of the shortest graph-spanning path, and the \target{girth} is the hop count of the shortest cycle in the graph~\cite{Dvorak2022}.

At the \scope{Subgraph} scope the target names apply to structures that do not incorporate all nodes, but include more than one node within the graph. \add{For smaller subgraphs, a generic name for a network pattern at the \scope{Subgraph} scope is \target{motif}~\cite{Shu2024}.} A subgraph that is a complete graph, namely where all pairs of nodes are connected by edges, is known as a \target{clique}. The related idea of a \target{cluster} is when the nodes have near-complete connectivity. A \target{cycle} is a loop: a path where the first and last vertices are equal. A \target{lacuna}, or structural hole~\cite{Lin2022}, is a cycle of 4 or more nodes without internal chords; the larger the cycle, the larger the lacuna. \add{A \target{star} covers a node that has two or more 1-degree singleton nodes connected to it~\cite{Sahu2020}. A \target{chain} is a linear sequence of connected nodes, each with a single edge between them.} 

\add{For larger subgraphs, the generic name for target at this scope is simply a \target{subgraph}.} A \target{geodesic} is the shortest path between two nodes~\cite{Newman2003}. The \target{core} is the most closely connected subset of nodes, and conversely the \target{periphery} is defined by exclusion to be the nodes not in the core which are not densely connected~\cite{Yoghourdjian2016}.

Finally, at the \scope{Constituent} scope we consider targets that only apply to a single \add{\target{node} or \target{edge}}. Both an \target{articulation} node and a \target{bridge} edge, when removed, disconnect the graph into two or more components~\cite{Lee2006}. This definition also relates to a target at the \scope{Subgraph} scope, a \target{bottleneck} region, a generalization and near-approximation of the bridge idea to a small subgraph. A \target{hub} is a node with a higher than average number of neighbors. \add{A \target{leaf} node is a singleton periphery node, only one edge connects it to the rest of the graph.} An \target{endpoint} is a diameter endpoint node in the graph's periphery; these always comes in pairs, and are found at opposite topological ends of the graph~\cite{Oddo2024}; while a \target{center} node is a node shared the graph's radius paths, a node found at the topological center of the graph.
\subsection{Action}

The \triplet{Action} dimension has six actions, which are not ordered. Many apply to more than one \triplet{Scope}; Fig.~\ref{TaskTaxonomy}-B summarizes these relationships.

\textbf{Group.} Asks \task{what groups are formed by considering similarity} through an unsupervised, bottom-up process that does not require a top-down categorization with named target structures. Applies to the \scope{Multiple} scope only.

\textbf{Categorize.} Asks \task{which kind of target is the graph?}, applicable to one graph at the \scope{Single} scope. At the \scope{Multiple} scope, this action can be applied to each of the graphs separately, using the targets at the \scope{Single} scope.

\textbf{Quantify.} Asks \task{how many targets does the graph have?} at the \scope{Constituent} or \scope{Subgraph} scopes, or \task{how many hops does the target span in the graph} at the \scope{Single} scope, or \task{how much are the graphs similar to each other} at the \scope{Pair} scope, or \task{how do all the graphs relate to each other in terms of similarity} at the \scope{Multiple} scope.

\textbf{Compare.} Asks \task{in what ways are the graphs similar/different?}, and applies to the \scope{Pair} scope only. \add{By similarity, we mean topological similarity of the network pattern~\cite{Shu2024}, namely any topological structure in a graph with some meaning; for example, whether the graph is connected or disconnected (Sec.~\ref{Sec:TaskTaxonomyScope}), or if there are same number of specific subgraph motifs like cliques, clusters, cycles, and so on (Sec.~\ref{Sec:TaskTaxonomyTargets}).}

\textbf{Classify.} Asks \task{does the graph contain any of the targets?} at the \scope{Subgraph} and \scope{Constituent} scope, or \task{is the graph an instance of the target?} at the \scope{Single} scope.

\textbf{Locate.} Asks \task{where are the targets situated within the graph?}, and applies them to the \scope{Subgraph} and \scope{Constituent} scopes. The location may involve proximity between targets of the same type, or between two different kinds of targets.

\vspace{-0.15cm}
\subsection{\add{Implications: Task Abstraction Model}}\label{Sec:TaskModelImplications}

\add{This taxonomy highlights tasks for which reasoning about topological structure  is directly relevant and useful (Sec.~\ref{Sec:DataAbstraction}). Fig.~\ref{TaskSets} shows how the set of labels-not-required tasks is nested within the larger set of network visualization tasks, aligning with our data abstraction model (Fig.~\ref{DataAbstraction}-A). The set of labels-required tasks is the complement set, or the set that remains when the subset of labels-not-required tasks is removed from the set of all network visualization tasks. Any task that requires lookup of a node or edge via data in the Label, Attribute, or Augmented layers is in the set of labels-required tasks.}


\add{Despite this Fig.~\ref{TaskSets} nesting, it is important to highlight that explicitly identifying the subset of labels-not-required tasks is useful: these tasks can benefit from the visual patterns surfaced by a purposefully designed invariant plot, rather than from visual patterns produced by network views traditionally tailored with labeled graph data considerations as a design requirement. Indeed, the fundamental rationale for this task taxonomy stems from the fact that a single unlabeled graph represents its entire isomorphism class identically (Fig.~\ref{DataAbstraction}-B, Sec.~\ref{Sec:DataAbstractionImplication}).}

\add{In contrast, with the labeled graphs behind traditional views, each graph member of its isomorphism class could be depicted differently. To illustrate, the degrees of freedom that result from the matrix seriation behind the Adjacency Matrix visual idiom yield different visual representations for graphs within the same isomorphism class. Similarly, the arbitrary spatial embedding of nodes as points in the Node-Link visual idiom allows for infinitely many representations of graphs within the same isomorphism class. In both of these cases, the design of the visual idioms implicitly relies on the availability of persistent labels to distinguish constituents from each other, even if those labels are not directly shown in a particular instantiation of the idiom
(more details in Sec.~\ref{Sec:VisualEncodings}).}



\section{Concrete Tasks}\label{Sec:ConcreteTasks}

We now connect the \add{\triplet{Target} + \triplet{Action} + \triplet{Scope}} triplets to concrete instances of tasks expressed as prose. \add{Then, in Sec.~\ref{Sec:ExampleAnalysis}, we analyze 6 different visual encodings (Fig.~\ref{VisualEncodings}) with respect to a list of 17 labels-not-required tasks (Fig.~\ref{ExampleAnalysis}).}

\vspace{-0.15cm}
\subsection{Previous Work Tasks}
 
We begin with examples of tasks articulated in previous work, showing the descriptive power of our task taxonomy. For example, the concrete task \task{How similar are these two graphs?} (T05) from Graph Thumbnails~\cite{Yoghourdjian2016} \add{two \target{graphs} as target} at the \scope{Pair} scope, with a \action{Quantify} action. Sharing a similar \add{\triplet{Target} + \triplet{Scope}} construction but different \triplet{Action}, the task \task{Are these two graphs structurally similar?} (T04) from Network Portrait literature~\cite{Bagrow2019} has a \action{Compare} action.

\vspace{-0.15cm}
\subsection{Generated Tasks}\label{Sec:GeneratedTasks}

We also articulate new tasks that have not appeared in previous task taxonomies, showing the generative power of our taxonomy. For example, \task{How many diameter endpoint nodes does the graph have?} (T17) \add{has the diameter \target{endpoint} node target at} the \scope{Constituent} scope, with a \scope{Quantify} action. \add{Another example,} \task{Where in the graph are the bridge edges?} (T14), also occurs at the \scope{Constituent} level, \add{but with a \target{bridge} edge target and} a \action{Locate} action.

\vspace{-0.15cm}
\subsection{Complex Tasks}\label{Sec:ComplexTasks}

Some concrete tasks are the combination of multiple simple tasks chained together~\cite{Brehmer2013}; we call these \emph{complex tasks}. One example is \task{Is the largest clique near the graph's periphery?}, which could be interpreted as a chained series of simpler tasks: it requires first identifying \target{cliques} within the graph through a \action{Classify} action \add{at the \scope{Subgraph} scope, followed by a} \action{Quantify} actions between those \target{cliques} to determine which of them is the largest, \add{and finally} a \action{Locate} action to \add{note the placement of the largest \target{clique}, also} at the \scope{Subgraph} scope. This last step maps to \task{Where in the graph are the cliques?} (T11) \add{in Fig.~\ref{ExampleAnalysis}, which considers} location with respect to the core-periphery \add{hierarchical} structure of the overall graph.

A compelling example of a complex task is chaining \scope{Subgraph} or \scope{Constituent} instantiated tasks to \action{Pair} and \action{Compare}, where the latter is a noteworthy \triplet{Scope} + \triplet{Action} singleton combination (Fig.~\ref{TaskTaxonomy}-B). Such chaining allows for the comparison of subgraph structures between different graphs, or even within the same graph if we consider the chained \scope{Pair} scope as duplicating the graph target. For example, \task{How similar is the largest hub in the sampled graph to the largest hub in the reference graph} is composed of \add{a \target{hub} target at the} \scope{Constituent} scope, and \scope{Quantify} action for the main abstract task, which is then chained to \add{two different \target{graphs} at the} \scope{Pair} scope and \action{Compare} action.

\vspace{-0.15cm}
\subsection{Ecological Validity of Tasks}\label{Sec:EcologicalValidity}

We can describe domain-language tasks found in the literature with the task abstraction provided by our taxonomy, providing evidence of both descriptive power and ecological validity. \add{The process of turning a concrete task into an abstract task starts with the \triplet{Target} noun and \triplet{Action} verb, which can be extracted from domain language in either order. The \triplet{Scope} can be inferred based on only the \triplet{Target} for simple tasks; while a mismatch between target and scope indicates a complex chained task that spans different scopes, as described in Sec.~\ref{Sec:ComplexTasks} above.}

The biological pathway literature states the task of \task{Identify potential feedback loops in gene regulation}~\cite{Murray2016} in domain language. We can abstract this task as having \add{topological \target{cycles} as target (feedback loop) with a \action{Classify} action, at the \scope{Subgraph} scope}.

In contrast, \task{Compare biological pathway to a pathway with the same functionality in a reference species}~\cite{Murray2016} \add{is a complex chained task because we have an action-scope mismatch between the \target{subgraph} as target (pathway) and the \action{Compare} action. Thus, the first step is a \target{subgraph} target with a \action{Locate} action at the \scope{Subgraph} scope, followed by a \target{graph} target with a \action{Compare} action at the \scope{Pair} scope.} 

The task \task{Which graph is the centroid (average) among a collection of graphs?}~\cite{Luppi2024}, to find the representative brain network among the graph outputs of an fMRI pipeline, comes from the neuroscience literature. \add{Since} multiple \target{graphs} are the target, \add{it maps to the \scope{Multiple} scope;} 
\add{the action is} \action{Quantify}. \add{This abstract triplet is the same as in \task{Rank graphs in order of similarity} (T03)} in Fig.~\ref{ExampleAnalysis}.

Another task articulated from a domain-specific perspective is \task{Monitoring for anomalous patterns in network traffic represented as a graph}~\cite{Sahu2020}, \add{in the context of preventing} cybercrimes. \add{It} can be abstracted with a \action{Classify} action that pertains to a domain-specific target which would specify an anomalous pattern in terms of some particular topological structure; \add{from this target, we know it occurs at the \scope{Subgraph} scope}. 

\vspace{-0.15cm}
\subsection{\add{Additional Concrete Tasks}}\label{Sec:AdditionalConcrete}

\add{We analyze a list of 30 tasks (26 chosen, 4 gleaned) from the Choose set of 15 papers (Fig.~\ref{PapersTable}), with a \add{\triplet{Target} + \triplet{Action} + \triplet{Scope}} triplet coding for each task. We now discuss three of these tasks; full table available in~\SuppTwo.}

\add{The task \task{choose a real network dataset from a line up with synthetic data}, chosen from GraphPrism~\cite{Kairam2012}, is a \target{graph} target, with a \action{Group} action, at the \scope{Multiple} scope. Task success instantiates as two groups: a singleton group that isolates the real-world network, and a remainder larger group of synthetic networks. Another simple task, \task{count the 2-connected components of a network}, chosen from Graph Thumbnails~\cite{Yoghourdjian2016}, is a \target{near-periphery} subgraph target, with a \action{Quantify} action, at the \scope{Subgraph} scope.}

\add{In contrast, the task \task{determine a graph's regularity, or whether it has a homogeneous degree variance}, gleaned from Network Portrait~\cite{Bagrow2008}, is a complex task: a \target{node degree} property as target with a \action{Quantify} action, followed by a \target{regular} degree-homogeneous graph type as target with a \action{Categorize} action; with both steps at the \scope{Single} scope. The first step quantifies differences in node degrees, and the second step confirms the \target{regular} graph type, which has no degree variance.}


\section{Example Analysis}\label{Sec:ExampleAnalysis}

We validate the descriptive and evaluative power of our taxonomy by using it to guide an example analysis in which we make educated guesses about the effort and success of a selection of 17 \add{labels-not-required tasks (Fig.~\ref{ExampleAnalysis})} applied to 6 visual \add{idioms (Fig.~\ref{VisualEncodings})}. \add{The visual patterns surfaced by these 6 different visual encodings yield visual similarities that may or may not map to an identical underlying network pattern~\cite{Shu2024}.} Our intent is to demonstrate that the taxonomy has good coverage of the breadth of tasks that pertain to unlabeled graphs, and enough precision to support the analysis of \add{invariant} visual \add{patterns}.

We explain the chosen encodings, elaborate on the analysis process, and summarize the analysis results.

\subsection{Visual \add{Idioms}}\label{Sec:VisualEncodings}

\begin{figure*}[!t]
\centering
\vspace{5pt}
\includegraphics[width=0.96\linewidth]{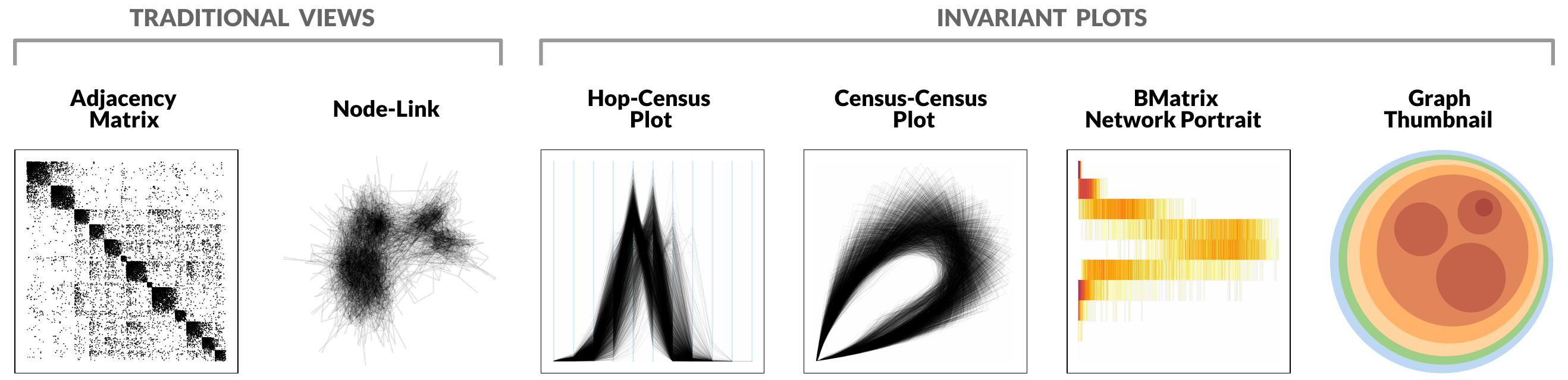}
\vspace{3pt}
\caption{The \add{6 static} visual \add{idioms} we consider for our assessment of \add{17 labels-not-required} tasks. \add{Here the} views share the same underlying graph topology: 'moreno-health' from the KONECT repository~\cite{Konect}, an empirical social network of 2539 nodes (people) and 10455 edges (friendships).}
\vspace{-8pt}
\label{VisualEncodings}
\end{figure*}

Fig.~\ref{VisualEncodings} shows the 6 static visual encodings that we analyze, ordered from high to low information fidelity; that is, the ability to reconstruct the full topology from the visual idiom. The first 2 are traditional network visual idioms and the latter 4 are \add{invariant plots}.


\textbf{Adjacency Matrix.} Among the visual idioms we consider, Adjacency Matrix is the one with the highest information fidelity: the entire graph topology can be reconstructed from the visual representation of the matrix data table. In the Adjacency Matrix visual idiom each node is assigned a unique row and column in a matrix, where non-zero cells at the intersection of two connected nodes represent their shared edge. When the represented graph topology is unlabeled \add{and undirected,} the visual patterns are always diagonally symmetric because there is no distinction between edge A-to-B and edge B-to-A. However, despite diagonal symmetry, surfacing salient visual patterns still depends on matrix seriation, namely the row and column permutation method, which is an algorithmic design decision. \add{Indeed, each matrix seriation maps to both a unique instantiation of shuffled node labels, and the visual pattern materialized as a result of these shuffled labels re-mapped to rows and columns of the matrix view.} In Fig.~\ref{VisualEncodings} we use generic Louvain communities to determine matrix seriation~\cite{Blondel2008}. Revealing network structure through the visual patterns that arise from matrix seriation is a non-trivial process that leads to varied outcomes~\cite{Behrisch2016}.

\textbf{Node-Link.} We also analyze a basic version of Node-Link layout, with edges represented by straight-line segments and nodes as equal-sized disks, monochromatic, and spatially arranged using a generic force-directed layout~\cite{Kobourov2013}. Although many more sophisticated \add{Node-Link} approaches have been proposed, most heuristics rely on Attributed values, which are unavailable in the Unlabeled context (Fig.~\ref{DataAbstraction}). Like matrix seriation in Adjacency Matrix, the spatial embedding of the node \add{points} in \add{the Node-Link view} is also a non-trivial process, often with extremely diverse visual \add{materialization} outcomes\add{, even for graph instantiations that are members of the same isomorphism class.}

\textbf{Hop-Census Plot.} In our previous work~\cite{Oddo2024}, we proposed a new line-based invariant plot we call Hop-Census, which encodes the Census data structure, which is composed of one vector of integers per node, and each vector is visually encoded as a polyline in a shared coordinate system. The vector integers represent counts of elementary constituents found during a traversal of the graph from a source node. The horizontal axis is bounded by the diameter of the graph. The vertical axis shows the counts of the elementary constituents, for example stub (half-edge) degree in the instantiation shown in Fig.~\ref{VisualEncodings}. The polyline for a given node captures the number of constituents at each hop: that is, the number of half-edges adjacent to the source node itself at the first hop, then the number of half-edges adjacent to the neighboring nodes at the second hop, and so on. Therefore, the Hop-Census plot is a type of parallel coordinate plot, where the horizontal ordering of the vertical axes is dictated by the data itself rather than being arbitrarily determined. Moreover, the extent of each vertical axis is uniform, providing an absolute coordinate system in which to visually embed the Census vector polylines. Thus, this plot shows a collective multi-polyline visual pattern that is itself invariant (i.e. always the same for all isomorphs of a graph).

\textbf{Census-Census Plot.} In our previous work~\cite{Oddo2024}, we also proposed Census-Census plots as another example of a line-based invariant plot. Census vectors can count different kinds of elementary constituents: stubs, nodes, and edges. These plots show two of these different kinds of Census vectors orthogonal to each other. The path traced by a single polyline follows a trajectory that reveals the differences between these different constituent counts, as the graph is traversed. In the instantiation shown in Fig.~\ref{VisualEncodings}, the horizontal axis is node degree and the vertical axis is stub degree. The length of a Census vector is always the same no matter which kind of constituent is counted, leading to an absolute shared coordinate system that affords this orthogonal juxtaposition. Like Hop-Census plots, the collective multi-trajectory visual pattern is invariant.

\textbf{BMatrix Network Portrait.} The Network Portrait idiom proposed by Bagrow et al.~\cite{Bagrow2008, Bagrow2019} is another invariant plot that uses a matrix heatmap to encode aggregated information about Unlabeled graphs, built upon a data structure of polylines equivalent to the Census vectors we discuss above. In contrast to the Hop-Census plots which directly draw polylines, the BMatrix Network Portrait heatmap idiom aggregates the information they contain into discretized colored cells that can reveal regions of density. The axes of the Network Portrait are transposed from the Census-based plots: the horizontal axis is generalized node degree, and the vertical axis is hop distances.

\textbf{Graph Thumbnails.} While Census plots and the Network Portrait rely on hop-based invariant data structures (Census vectors and BMatrix, respectively), the Graph Thumbnails idiom~\cite{Yoghourdjian2016} relies on a different invariant data structure: the k-core decomposition of the graph. The definition of a k-core subgraph is that the value k is equal to the minimum degree of all nodes within it, which effectively decomposes the graph into nested subgraphs; every node within a k-core subgraph has at least degree k. Graph Thumbnails uses concentric bubbles to aggregate nodes of the same k-core subgraph into a single bubble, and then color encodes k-core hierarchy with a heatmap from outermost peripheral 1-core (blue), to inner 2-core (green), to 3-core (bright orange), and so on to the innermost cores (increasingly darker orange hues). The nested structure of Graph Thumbnails arises from a trade-off of information expressiveness that maximizes the salience of the graph's k-core decomposition to highlight core-periphery structure, while letting go of information fidelity through the choice of concentric bubble visual encoding: the original topology is aggregated and impossible to reconstruct from the visual idiom.

\vspace{-0.1cm}
\subsection{Analysis Dimensions}\label{Sec:AnalysisDimensions}

For each possible combination of abstract \add{\triplet{Action} + \triplet{Scope}} enumerated in the taxonomy (Sec.~\ref{Sec:TaskTaxonomy}), we sampled one or two \triplet{Target} possibilities to select one or two concrete tasks. The resulting list of tasks provides evidence of the descriptive power of the taxonomy in terms of coverage. We see that the \add{\triplet{Scope}} dimension leads to good coverage across ranges of analysis granularity; as we discuss in Sec.~\ref{Sec:RelatedWork}, previous taxonomies do not provide adequate coverage of the \add{\scope{Pair}} and \add{\scope{Multiple}} levels. This triplet-based analysis provides broad coverage of the full suite of unlabeled graph considerations; previous task taxonomies that focus on labeled graphs do not provide a way to systematically enumerate \add{labels-not-required} tasks.

In some cases our choice of which task to sample for the triplet was guided by tasks that are particularly well handled by one of the invariant plots, to illuminate the interesting possibilities in these under-studied invariant techniques. We then made educated guesses about the performance of each visual encoding according to two criteria: the Effort required from the viewer to carry out the task, and the likelihood of Success across a range of datasets with different characteristics.

The first author carried out the first round of assessment through inspection of each of the six visual encodings across an eclectic collection of 81 unlabeled graphs. This benchmark dataset was already demonstrated on \add{5 visual idioms
} in our own previous work~\cite{Oddo2024}\add{. We add Graph Thumbnails as a 6th idiom because this technique is intentionally designed to surface visual patterns from the invariant core-periphery hierarchical decomposition of a network, an invariant descriptor not addressed by the other visual idioms}. \add{These} 9-by-9 collages showcase the 81 networks from our eclectic benchmark at once; one collage per each of the \add{6 visual idioms}~\add{(\SuppFive)}. All authors discussed the assessment, handling disagreements through iterative rounds of analysis until convergence.

To avoid a coarse binarized pass-or-fail assessment, we iterated and settled into 4 bins so our results capture more nuance of our educated guess consensus on visual encoding performance. Our analysis dimensions Effort and Success helped us better capture our different points of view simultaneously and arrive at consensus. For the Effort dimension (i.e. our estimate of the cognitive load demanded from human viewers) our bins are Easy, Middling, Difficult, and Impossible. For the Success dimension (i.e. our estimate of the likelihood of task completion) our bins are Always, Sometimes, Rarely, and Never. This qualitative assessment is both preliminary and subjective, and should be treated as a starting point rather than a definitive statement of facts: our intent is to spur future work that uses the taxonomy to conduct empirical studies.

We also consider a Scale dimension (i.e. the cardinality of the graph) with two sizes of graph data: Small-scale (\texttt{<}100 nodes) and Large-scale (\texttt{>}1000 nodes), a size distinction following survey results~\cite{DiBartolomeo2024}. Our goal was to roughly divide visual encodings according to contexts where each works well, against scenarios where they begin to fail. For example, large graphs would have so much clutter when using basic force-directed placement that they would appear to be \emph{hairballs}~\cite{Munzner2014}. We consider relatively sparse graphs that are common in visualization contexts, where there are no more than three times the number of edges vs nodes~\cite{Melancon2006}.

\vspace{-0.1cm}

\subsection{Analysis Results}

\begin{figure*}[!t]
\vspace{2pt}
\centering
\includegraphics[width=0.9\linewidth]{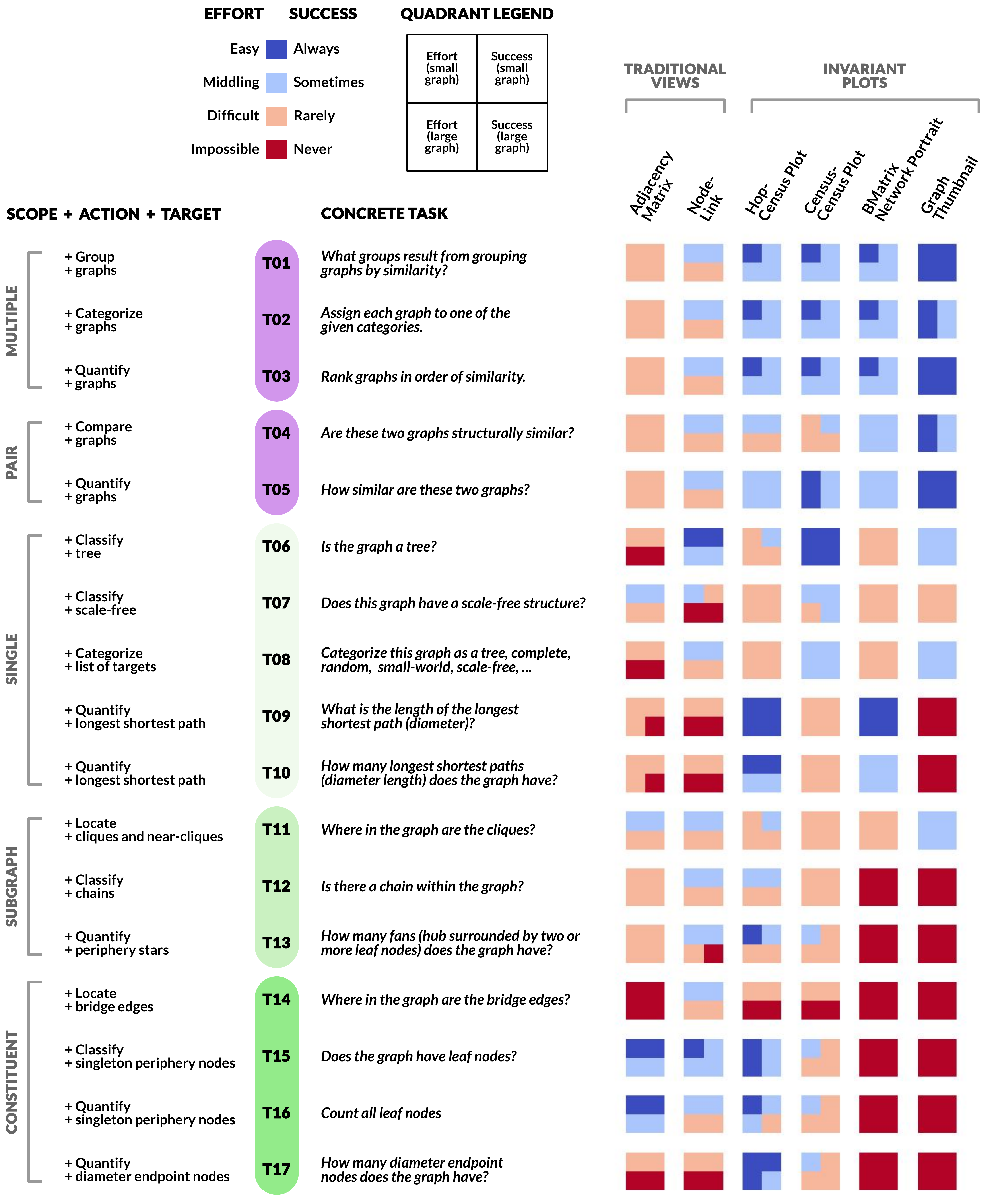}
\vspace{-6pt}
\caption{We illustrate the descriptive and evaluative power of our taxonomy \add{with 17 labels-not-required tasks} covering all possible \add{\triplet{Action} + \triplet{Scope}} combinations with one or two sampled from \triplet{Target} (rows) to consider with respect to 6 \add{visual idioms} (columns). At each task and visual encoding intersection, the table shows our assessment of user performance vs.~data scale with a quadrant heatmap: top left for task effort on Small graphs, top right for task success on Small graphs, bottom left for task effort on Large graphs, and bottom right for task success on Large graphs. Task effort and success are indicated with a diverging blue-red colormap mapped to 4 ranked assessment levels: Easy, Middling, Difficult, and Impossible (for task effort); and Always, Sometimes, Rarely, and Never (for task success).}
\vspace{-8pt}
\label{ExampleAnalysis}
\end{figure*}

Fig.~\ref{ExampleAnalysis} shows the summarized analysis results; \add{full spreadsheet} available in~\add{\SuppFour}, \add{which includes a} brief rationale for each cell.


The overall structure visible in our assessment summary shows the impact between the task \triplet{Scope} (from broad-scope tasks at the top in purple, to narrow-scope tasks at the bottom in green) and visual encoding information fidelity (from high on the left to low on the right). There is good performance (blue) on the upper right at the \scope{Multiple} and \scope{Pair} scopes, variable performance at the \scope{Single} scope, and mostly poor performance (red) on the lower right at the narrower scopes of \scope{Subgraph} and \scope{Constituent}. Conversely, the left side fares better at the narrower scopes, and worse at the broader ones. 

We consider that the Adjacency Matrix idiom (first column) has a low performance for the multi-graph scopes for two reasons: first, we expect users to have perceptual and cognitive difficulties reading the Adjacency Matrix visual pattern~\cite{Abdelaal2022}; and second, because the visual pattern itself relies on a non-trivial matrix seriation choice~\cite{Shu2024}. In the unlabeled graph context, where there are no persistent labels, there is no way to find correspondences between nodes across graphs, so matrix seriation is unlikely to reveal similarities between them that could be revealed through visual patterns (T01-T05). The Adjacency Matrix idiom excels at a few tasks at the \scope{Constituent} level, namely the \action{Quantify} (T16) and \action{Classify} (T15) tasks for \add{\target{leaf}} node targets, where counting and finding matrix rows that have only one cell is easy for small graphs and middling for large graphs. The well-known challenge of Adjacency Matrix at handling tasks that involve tracing topological paths is apparent for many of the narrow-scope targets, including \target{bridges} (T14), \target{chains} (T12), and \target{diameters} (T09, T10, T17). 

The Node-Link idiom (second column) performance is frequently affected by scale, with many quadrants that differ between the top and bottom, showing tasks that are handled fairly well at the small scale but pose difficulties at the large scale. The performance difference is due to extensive overlapping and occlusion, yielding hairballs that are visually indistinguishable despite topological differences in the underlying data. 

The BMatrix Network Portrait and Graph Thumbnails idioms (fifth and sixth columns) do poorly on the narrow-scope tasks at the bottom of the table (T12-T17): the Small-scale \scope{Subgraph} and \scope{Constituent} tasks are almost all impossible to perform, because these visual idioms aggregate away node-specific information. However, they do well at the broad-scope \scope{Multiple} and \scope{Pair} tasks. The aggressive aggregation of Graph Thumbnails yields clear views of high-level topological structure that are robust against changes of scale. The BMatrix Network Portrait approach has a more moderate approach to aggregation and concomitantly a more moderate resilience to scale changes. 
 
The Hop-Census Plot and Census-Census Plot invariant plot idioms (third and fourth columns) also do best at the multi-graph \scope{Multiple} and \scope{Pair} broad-scopes, in part due to their shared frame of reference affordance~\cite{Oddo2024}. They also have some resilience to scale changes in those scopes. The Hop-Census idiom has good performance at tasks involving \target{diameter} targets at the \scope{Single} scope (T09, T10) as does the BMatrix Network Portrait, as these idioms directly visually encode the relevant information in a way that can be easily counted for \action{Quantify} actions and spotted for \action{Classify} actions. Hop-Census is unique among the invariant idioms in also performing well at many tasks in the \scope{Subgraph} and \scope{Constituent} scopes, including \target{diameter} (T17) and \add{\target{leaf} node} (T13, T15, T16) targets. It also performs reasonably well for the \target{chain} target at the \scope{Subgraph} scope, which can sometimes be spotted with middling effort because topological chains have a distinct constant-value polyline across parallel hop axes. The same performance applies to Node-Link views for this task, where chains may be visible in the layout.

Some \scope{Single} scope tasks have great variability. For example, \task{Is the graph a tree?} (T06) is nearly impossible with the Adjacency Matrix view, while trivial with an appropriately laid out Node-Link view. For the invariant idioms, it is difficult for the Hop-Census and BMatrix idioms, but easy to spot the characteristic patterns in Census-Census: a crisp positive-slope line.

For \task{Does this graph have scale-free structure} (T07), the Adjacency Matrix view can sometimes handle small graphs with middling effort using the appropriate matrix seriation, where the scale-free pattern can be made salient by sorting according to number of neighbors and looking for an exponential decay pattern; with large graphs, the effort is large and the success is rare as the matrix grows large and the pixels grow small. Notably, the challenges in seriation at the broad-scope does not apply here, because there is no need to find node correspondences between graphs. The Node-Link view would require middling effort for small graphs, where the scale-free pattern is more likely to be salient, but would not work at all in the large cases. The Census-Census plots can sometimes verify scale-free structure for small graphs with middling effort, although they more easily rule out when the topology is not scale-free; for large graphs, occlusion leads to difficult status, although sometimes scale-free patterns can still surface. The aggregate structure of the BMatrix Network Portrait and Graph Thumbnails views leads to difficult effort and rare success at both scales. 

Another interesting concrete task is \task{Where in the graph are the cliques?} (T11). Both traditional views have reasonable performance at the Small-scale: in the Adjacency Matrix clusters can be made salient with the right permutation, and in the Node-Link view dense regions are usually distinguishable from sparse ones. The Hop-Census plot sometimes achieves success, but with a high level of effort: if the polylines of the cluster nodes span most of the hop axis, then the cluster is more peripheral than central. The Census-Census and BMatrix Network Portrait idioms do not perform well at this task, but the Graph Thumbnails idiom may yield good results: the color of inner core bubbles provides the answer, which is highly robust against scale. 

Our assessment summary (Fig.~\ref{ExampleAnalysis}) also shows that the effect of \triplet{Action} on idiom performance is often small, for tasks which share both \triplet{Scope} and \triplet{Target}. In some concrete cases, the \action{Classify} task is easier than the \action{Quantify} task, as with noting the existence of a \add{\target{leaf}} node target \add{\task{Does the graph have leaf nodes?}} (T15) versus counting the number of instances \add{\task{Count all leaf nodes}} (T16).


\vspace{-0.1cm}

\section{Discussion}\label{Sec:Discussion}

We discuss the utility of our task taxonomy, the limitations of our study, and future work directions.

\vspace{-0.2cm}

\subsection{Utility}

Taxonomies can be judged according to their descriptive, evaluative, and generative power~\cite{BeaudouinLafon2004}. We provide evidence of the descriptive power of our taxonomy by connecting the fully abstracted task triplets to concrete tasks proposed in the previous work. Evaluative power was one of our major motivations in creating it: we wanted to help future researchers in designing user studies, when they are deciding on what tasks would be appropriate. Our preliminary assessment \add{across different} visual idioms with the taxonomy does provide some evidence of its evaluative and descriptive power. We also think the taxonomy has great promise in terms of generative power. We have generated some examples of concrete tasks not previously articulated in task taxonomies considering the combinatorial possibilities of abstract triplets.

We developed this taxonomy for the pragmatic reason that existing invariant plot methods cannot be evaluated without it. However, conversely, we hope that the existence of this task taxonomy will spur future work. In particular, Fig.~\ref{ExampleAnalysis} strongly indicates the potential benefit of future technique-oriented work to address some of the regions where existing methods fall short. \add{In Sec.~\ref{Sec:AnalysisDimensions} we propose the analysis dimensions of Effort, Success, and Scale as a guide for future empirical studies that could incorporate them directly into experimental design.}

\vspace{-0.2cm}

\subsection{Limitations and Future Work}

The most obvious limitation of our work is that our assessment of visual idioms is highly subjective and should be considered an educated guess. We intend our analysis to illustrate the evaluative power of our task taxonomy, where we primarily consider whether the visual encodings make the particular target visually salient, or conversely whether the visually salient features in the encodings correspond to the targets of interest. We drew upon the knowledge of all authors, but in many cases we were well aware that more research is needed to further explore these questions; especially for the less well studied invariant plots. We call for future empirical work through human-subjects studies that could serve to either confirm or modify \add{our} conjectures.

Moreover, a different concrete task than the one that we selected for the abstract example may give very different results. Each abstract \add{\triplet{Target} + \triplet{Action} + \triplet{Scope}} triplet gives rise to many concrete tasks, and an Effort and Success analysis of them may be quite different. The ones that we selected to analyze cannot possibly tell a complete story about the entire class of possibilities. We also encourage future empirical work to assess more concrete tasks, to have better coverage of the full suite of possibilities.

We developed our data abstraction (Sec.~\ref{Sec:DataAbstraction}) as a filtering tool to analyze tasks. Future research may also use this model to investigate graph-theoretic and algorithmic properties of unlabeled graphs in relation to semantic richness, with applications that can extend beyond visualization given the ubiquity of data with connected topology~\cite{Sahu2020}.

Finally, we see a need for even more language to describe directions and locations in purely topological terms, without any references to a particular geometric layout of nodes and edges. In this work we propose the \target{core-periphery} \add{hierarchical decomposition} and \target{bottleneck} regions as helpful language for the \scope{Subgraph} scope; we call for more identification of \add{such topological wayfinding} targets in future work.



\section{Conclusion}\label{Sec:Conclusion}

We propose a taxonomy for \add{labels-not-required network visualization} tasks organized by \add{\triplet{Target} + \triplet{Action} + \triplet{Scope}} triplets. We validate the descriptive, generative, and evaluative power of this taxonomy by considering 6 visual \add{idioms} (2 traditional views and 4 invariant plots) across one task sampled from each triplet. We then perform an educated guess assessment of encoding usability for each task, as an example of how user study design could make use of our taxonomy. We encourage formal human-subject studies using the taxonomy to verify the utility of \add{invariant plots} that lack quantitative and qualitative evaluations, such as Network Portrait, Graph Thumbnails, and Census plots.


\vspace{-0.2cm}
\section*{Acknowledgements}

The development of this paper took place between October 2024 and \add{January 2026} at the UBC Point Grey campus and in the City of Vancouver, which sit on the traditional, ancestral, unceded territory of the Musqueam, Squamish, and Tsleil-Waututh First Nations. This work was supported in part by NSERC DG RGPIN-2024-06401. We thank Steve Kasica, Firas Moosvi, Francis Nguyen, and Mara Solen for their manuscript reviews.

\vspace{-0.2cm}


\bibliographystyle{IEEEtran}
\bibliography{IEEEabrv, REVREV_references}

\vskip -2\baselineskip plus -1fil

\begin{IEEEbiography}[{\includegraphics[width=1in,height=1.25in,clip,keepaspectratio]{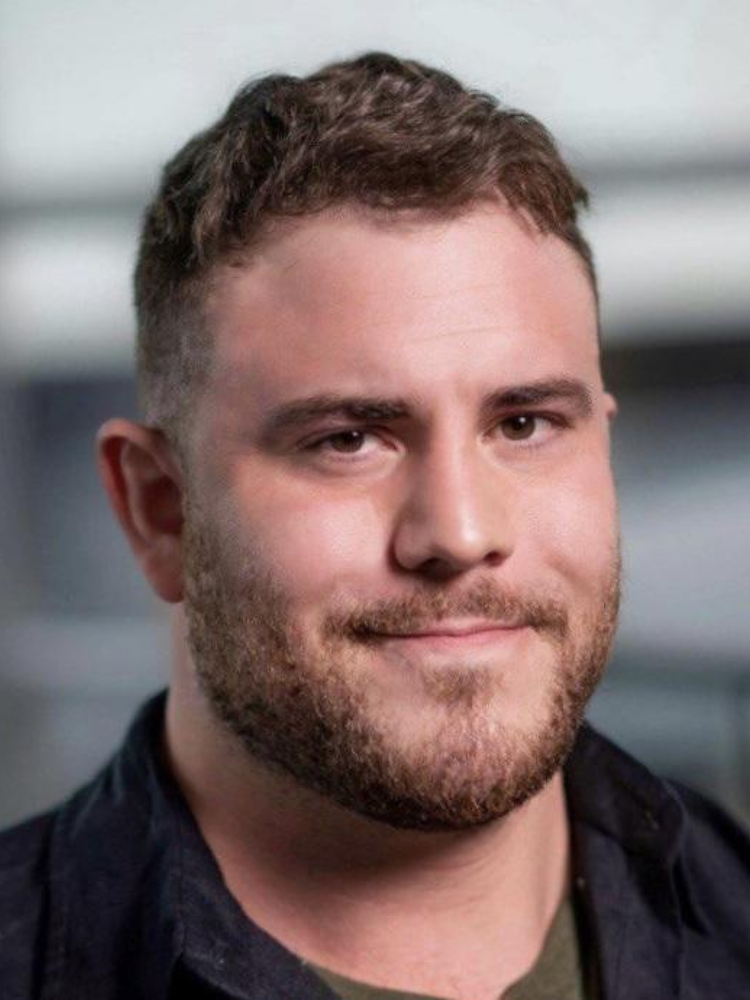}}]{Matt I. B. Oddo} Matías Ignacio Bofarull Oddó (Matt Oddo) is a Chilean-Canadian PhD candidate studying Information Visualization in the Department of Computer Science at the University of British Columbia since 2022. He received the BSc degree in biology from UBC Okanagan and the MSc in earth science from UBC EOAS. Previously he was a MITACS researcher at the Institute for the Oceans and Fisheries and a data scientist at Lucent Biosciences.
\end{IEEEbiography}

\vspace{-0.2cm}

\vskip -2\baselineskip plus -1fil

\begin{IEEEbiography}[{\includegraphics[width=1in,height=1.25in,clip,keepaspectratio]{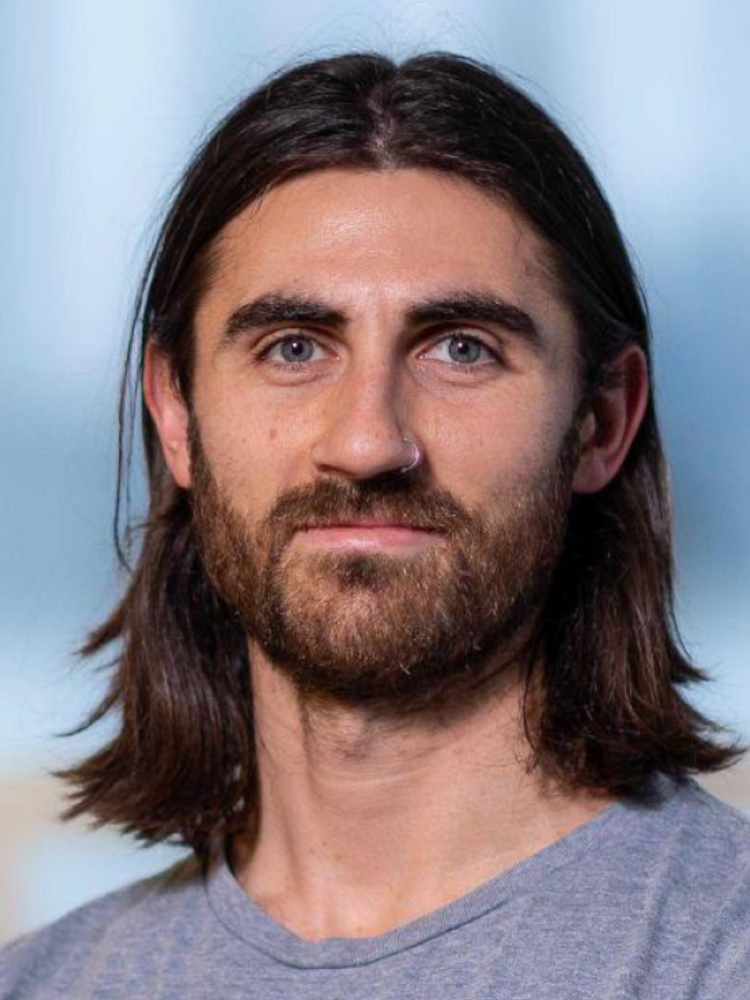}}]{Ryan Smith} is a PhD candidate and MITACS researcher studying Information Visualization in the Department of Computer Science at the University of British Columbia since 2022. He received the BA degree in Psychology from Marist College and the MA in Psychology from New York University. His research focuses on how patient-centric visualization can improve healthcare.
\end{IEEEbiography}

\vskip -2\baselineskip plus -1fil

\begin{IEEEbiography}[{\includegraphics[width=1in,height=1.25in,clip,keepaspectratio]{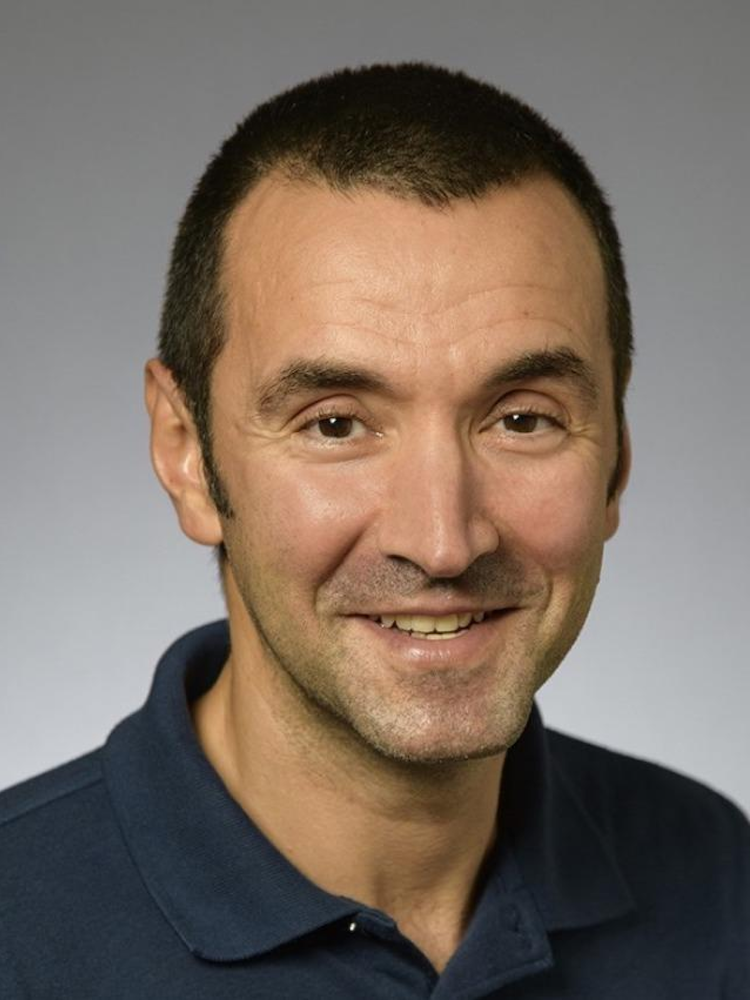}}]{Stephen Kobourov} is a Professor of Computer Science at the Technical University of Munich. He received the BS degrees in Mathematics and Computer Science from Dartmouth College and the MS and PhD degrees from Johns Hopkins University. His research interests include information visualization, graph theory, and geometric algorithms.
\end{IEEEbiography}

\vskip -2\baselineskip plus -1fil

\begin{IEEEbiography}[{\includegraphics[width=1in,height=1.25in,clip,keepaspectratio]{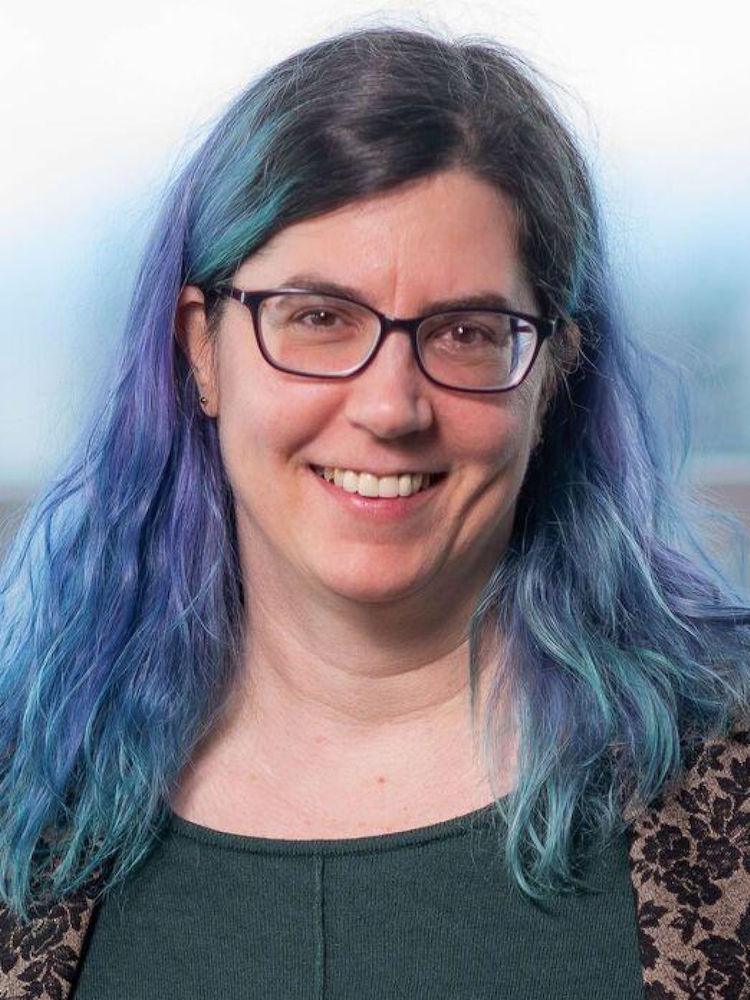}}]{Tamara Munzner}
(IEEE Fellow) is a Professor with the University of British Columbia. She received the PhD degree from Stanford. She has worked on visualization projects in a broad range of application domains from genomics to journalism. Her book Visualization Analysis and Design is heavily used worldwide, and she was the recipient of the IEEE VGTC Visualization Technical Achievement Award.
\end{IEEEbiography}

\end{document}